\documentclass[pra,twocolumn,nofootinbib,letter]{revtex4}
\usepackage[english]{babel}
\usepackage{graphicx}
\usepackage{bm,amsmath,amssymb}
\usepackage{dcolumn,longtable}
\usepackage{natbib}
\usepackage[colorlinks]{hyperref}
\usepackage{float}
\usepackage{wrapfig}

\newcommand{\pten}[2]{\ensuremath \left( \begin{array}{ccc}
#1 \\[0.3cm]
#2 \\ 
\end{array} \right) }

\begin{document}

\title{Dependence of atomic parity-violation effects on neutron skins and new physics}

\author{A. V. Viatkina$^{1}$}
\author{D. Antypas$^{1}$}
\author{M. G. Kozlov$^{2,3}$}
\author{D. Budker$^{1,4}$}
\author{V. V. Flambaum$^{1,5}$}
\affiliation{$^1$Helmholtz Institute, Johannes Gutenberg-Universit\"{a}t Mainz, 55099 Mainz, Germany}
\affiliation{$^2$Petersburg Nuclear Physics Institute of NRC "Kurchatov center", Gatchina, Leningrad District 188300, Russia}
\affiliation{$^3$St.\,Petersburg Electrotechnical University LETI, Prof.\,Popov Str. 5, 197376 St.\,Petersburg, Russia}
\affiliation{$^4$Department of Physics, University of California, Berkeley, California 94720-7300, USA}
\affiliation{$^5$School of Physics, University of New South Wales, Sydney 2052, Australia}
\date{\today}

\begin{abstract}
We estimate the relative contribution of nuclear structure and new physics couplings to the parity non-conserving spin-independent effects in atomic systems, for both single isotopes and 
isotopic ratios. General expressions are presented to assess the sensitivity of isotopic ratios to neutron skins and to couplings beyond standard model at tree level. The specific coefficients for these contributions are calculated assuming Fermi distribution for proton and neutron nuclear densities for isotopes of Cs, Ba, Sm, Dy, Yb, Pb, Fr, and Ra. The present work aims to provide a guide to the choice of the best isotopes and pairs of isotopes for conducting atomic PNC measurements.
\end{abstract}

\maketitle

\section{Introduction}\label{sec_intro}
Parity-nonconserving (PNC) effects observed in low-energy atomic experiments can provide a precise test of the standard model (SM) and serve as probes of its possible extensions. For a range of new-physics scenarios, notably for those which manifest themselves at lower energies, atomic tests can give complementary and sometimes better probe compared to experiments at high-energy colliders \cite{safronova_search_2018}.  Currently, a number of experiments are being conducted, with the aim to measure the PNC effects in isotopic chains, for example Yb experiment in Mainz \cite{antypas_isotopic_2019} and Fr experiment at TRIUMF \cite{zhang_efficient_2016}. 

Parity violating weak interaction between electron and quarks in a nucleus is mediated by the heavy Z$^0$-boson ($M_Z\approx 91.2$~GeV), therefore on atomic and even nuclear energy scales it can be considered point-like. This interaction is given by the product of axial and vector neutral currents. One can neglect internal nucleon structure and combine individual quarks into protons and neutrons \cite{pollock_atomic_1992}, in which case the PNC part of the Hamiltonian for neutral currents is written as:
\begin{gather}
\hat{h}_\mathrm{PNC}=\hat{h}_{SI}+\hat{h}_{SD} \nonumber \\
=\frac{G_f}{\sqrt{2}}\sum_{B}[C_{1B}\overline{e}\gamma_\mu\gamma_5 e \overline{B}\gamma_\mu B
+ C_{2B}\overline{e}\gamma_\mu e \overline{B}\gamma_\mu \gamma_5 B ]\ ,\label{eq:ham}
\end{gather}
where the two terms represent the nuclear spin independent (SI) and dependent (SD) parts, respectively. The operators $B$ and $e$ are nucleon and electron field operators, $B$ runs over all protons ($p$) and neutrons ($n$) in the nucleus, $\gamma_\mu$ ($\mu=0,1,2,3$) and $\gamma_5$ are Dirac matrices and 
\begin{equation}
G_f\approx 1.16638\times 10^{-5}\ (\hbar c)^3\mathrm{GeV}^{-2} 
\end{equation}
is the Fermi weak-interaction constant. 
With $\theta_W$ being the Weinberg angle ($\sin^2\theta_W\approx 0.23$) and nucleon axial charge being $g_A\approx 1.26$, the coefficients from \eqref{eq:ham}, calculated in the lowest tree-level approximation of SM, are
\begin{align}
C_{1p}&=\frac{1}{2}(1-4\sin^2\theta_W)\approx 0.04,  \label{eq:c1p}\\ 
C_{1n}&=-\frac{1}{2},  \label{eq:c1n}\\ 
C_{2p}&=-C_{2n}=\frac{1}{2}(1-4\sin^2\theta_W)g_A\approx 0.05\ .  \label{eq:c2p_n}
\end{align}

Spin-independent $\hat{h}_{SI}$ arises from the coupling of electron axial current and nucleon vector current. On the other hand, $\hat{h}_{SD}$ comes from the coupling of electron vector current and nucleon axial current. All nucleons contribute coherently to $\hat{h}_{SI}$, but protons considerably less than neutrons (since $C_{1p}\ll C_{1n}$). In SD part only the valence nucleon contributes, moreover, both $C_{2p}$ and $C_{2n}$ are small. The $\hat{h}_{SD}$ contribution is thus typically three orders of magnitude smaller than that of $\hat{h}_{SI}$. It should be noted that there exist other spin-dependent PNC contributions, arising for example from nuclear anapole moment, but they are usually still small in comparison to the first term of \eqref{eq:ham} \cite{flambaum_p-odd_1980,ginges_violations_2004}. 

In this work we focus on $\hat{h}_{SI}$. Treating nucleons non-relativistically, one can rewrite it as an effective single-electron operator
\begin{equation}
\hat{h}_{SI}=\frac{G_f}{\sqrt{2}}\left[C_{1p}Z\rho_p(r)+C_{1n} N\rho_n(r)\right]\gamma_5 \ ,\label{eq:ham_pn}
\end{equation}
where $\rho_p$ and $\rho_n$ signify proton and neutron densities as functions of a radial variable $r$, both normalized to unity. Most often, the difference of $\rho_p$ and $\rho_n$ is neglected and \eqref{eq:ham_pn} turns into
\begin{gather}
\hat{h}_{SI,0}=\frac{G_f}{2\sqrt{2}}Q_W\rho(r)\gamma_5\ \label{eq:h_SI_0},\\
Q_W\equiv 2C_{1p}Z +2C_{1n} N \ ,\label{eq:Q_W_traditional}\\
\rho_p(r)=\rho_n(r)\equiv\rho(r)\ .
\end{gather}
Here $Q_W$ is the weak charge of the nucleus. Within tree-level of SM, the weak charge is given by
\begin{equation}
Q_W=Z(1-4\sin^2\theta_W)-N\ .\label{eq:QW_tree}
\end{equation}
The more precise SM value of $Q_W$ with radiative corrections included, accurate at the $0.1\%$ level, is \cite{tanabashi_review_2018}:
\begin{equation}
Q_W\approx -0.989 N+0.071 Z\ .\label{eq:QW_radiative}
\end{equation}
The values of \eqref{eq:QW_tree} and \eqref{eq:QW_radiative} differ typically by $0.5\%$, for the isotopes investigated in Section \ref{sec:evaluation}.\footnote{Radiative corrections from Eq.\,\eqref{eq:QW_radiative} should be added to the final results for $\tilde{Q}$.}

For the purposes of this work, it is necessary to account for the difference between proton $\rho_p$ and neutron $\rho_n$ nuclear densities. Let us return to the effective Hamiltonian \eqref{eq:ham_pn}. This operator enters PNC atomic observables through a matrix element between two atomic states with relativistic wave functions $\psi_j$ and $\psi_i$:
\begin{align}
\mathcal{M}=&\langle j|\hat{h}_{SI}| i\rangle\nonumber\\
=& \frac{G_f}{2\sqrt{2}}\Big[2 C_{1p}Z\int\rho_p(r)\psi^\dagger_j\gamma_5\psi_i d^3r\nonumber\\
&+2C_{1n}N\int\rho_n(r)\psi^\dagger_j\gamma_5\psi_i d^3r\Big].\label{eq:ME1}
\end{align}
To simplify further calculations, we consider only the single-electron $\psi$ of a valence electron. Due to the point-like nature of the weak interaction, the matrix element \eqref{eq:ME1} is the largest between $s$ and $p_{1/2}$ states. Using the Dirac wave functions for these orbitals inside a uniformly charged spherical nucleus and integrating over angles (see Appendix), one can show that
\begin{equation}
\int\! d\phi\int\! d\theta\ \psi_{p_{1/2}}^\dagger\gamma_5\psi_s =\mathcal{A}_{ps} \mathcal{N} f_{ps}(r)\ ,\label{eq:f}
\end{equation}
and
\begin{equation}
\mathcal{M}=\frac{G_f}{2\sqrt{2}} \mathcal{A}_{ps}\mathcal{N}\tilde{Q}_{W,ps}\ ,\label{eq:M}
\end{equation}
where $\mathcal{A}_{ps}$ is the atomic factor which has no dependence on nuclear parameters\footnote{In case one wants to account for it, a small nuclear-parameter-dependent contribution of an isotope shift to atomic wave functions should be included in the normalization factor $\mathcal{N}$.}, $\mathcal{N}$ is a normalization factor dependent on the nuclear charge radius, $f_{ps}(r)$ is the spatial variation of electron wave function density over the nucleus with $f_{ps}(0)=1$, and $\tilde{Q}_{W,ps}$ is an effective nuclear weak charge, which quantifies the coupling strength between $s$ and $p_{1/2}$ electrons. For simplicity, let us label $\tilde{Q}_{W,ps}\equiv\tilde{Q}$ and $f_{ps}(r')\equiv f(r')$. Within the limits of SM and at the tree-level approximation, one writes:
\begin{align}
\tilde{Q}&=Zq_p(1-4\sin^2\theta_W)-Nq_n ,\label{eq:q_w}\\
q_p&=\int\rho_p(r')f(r')dr' ,\label{eq:q_p}\\
q_n&=\int\rho_n(r')f(r')dr' .\label{eq:q_n}
\end{align}
Here $\rho_p$ and $\rho_n$ are normalized to unity. Throughout the present work, we assume proton density equal to charge density $\rho_p= \rho_\mathrm{charge}$. \footnote{A small difference is examined in \cite{ong_effect_2010}.}

It is useful to introduce an equivalent radius $R$ as a substitute for the rms radius $r_\mathrm{rms}=\sqrt{\langle r^2\rangle}$. In the case of a sharp-edge uniform spherical nucleus (SE model), from the definition of $\langle r^2\rangle$, the two radii are related as: 
\begin{equation}
r_\mathrm{rms}^2\equiv \langle r^2 \rangle=\frac{4\pi\int_0^{R} r'^2\cdot r'^2dr'}{4\pi\int_0^{R}r'^2dr'}=\frac{3}{5}R^2\ .\label{eq:R_SE}
\end{equation}
It should be noted that the expression of $R$ through $r_\mathrm{rms}$ varies according to the nuclear density model used.

As follows from expressions for $s$ and $p_{1/2}$ wave functions presented in Appendix, the normalization factor $\mathcal{N}$ in \eqref{eq:M} can be expressed through the proton distribution equivalent radius:
\begin{equation}
\mathcal{N}=\left(\frac{2 Z R_p}{a_B}\right)^{2\gamma -2} .
\end{equation}
Here $a_B$ is Bohr radius and $\gamma =\sqrt{1-(\alpha Z)^2}$, with $\alpha$ being fine structure constant and $Z$ the atomic number.

Some classes of interactions beyond SM would contribute to the observed nuclear weak charge $Q_W$ \cite{ramsey-musolf_low-energy_1999} and, correspondingly, to the effective nuclear weak charge $\tilde{Q}$:
\begin{equation}
\tilde{Q}_\mathrm{new} = \tilde{Q}+\Delta \tilde{Q}\ .\label{eq:dQ_correction}
\end{equation}
An observation of $\Delta\tilde{Q}$ would be a hint for beyond SM interactions. 

Uncertainties in the knowledge of the neutron distribution $\rho_n$, however, may hinder the extraction of $\Delta\tilde{Q}$ for neutron-rich nuclei. The problem in our case would be the poorly known radius of neutron distribution for most elements. The difference between radii of neutron and proton distributions is often referred to as \textit{neutron skin}. For example, it was measured in $^{208}$Pb by PREX collaboration to be $0.33^{+0.16}_{-0.18}$~fm \cite{prex_collaboration_measurement_2012}.

In \cite{wood_measurement_1997}, the PNC transition amplitude between the 6s and 7s states of neutral $^{133}$Cs was measured with an accuracy of 0.35\%. There is only one electron above closed shells in Cs, which makes precise atomic calculations feasible and the electronic part $\mathcal{A}_{ps}$ can be accounted for by numerical calculations without acquiring significant errors \cite{dzuba_high-precision_2002,porsev_precision_2009,dzuba_revisiting_2012}. Taking into account neutron-distribution corrections \cite{derevianko_correlated_2001}, $Q_W$ of Cs was found to be in agreement with SM, leading to strong constraints on possible new-physics scenarios.

In heavy atoms, the PNC effect should be enhanced, since the atomic PNC observable related to \eqref{eq:M} grows faster than $Z^3$ \cite{bouchiat_i._1974}, but the interpretation of single-isotope measurement requires precise evaluation of $\mathcal{A}_{ps}$, which, for atoms with multiple valence electrons, is far from trivial. It was proposed \cite{dzuba_enhancement_1986} to measure ratios of PNC matrix elements in different isotopes of the same atom, thereby canceling out the atomic part:
\begin{equation}
\mathcal{R}=\frac{\mathcal{M}'}{\mathcal{M}}=\left(\frac{R'_p}{R_p}\right)^{2\gamma-2}\frac{\tilde{Q}'}{\tilde{Q}}. \label{eq:ratio}
\end{equation}
Here $\mathcal{M}$ and $\mathcal{M}'$ correspond to the matrix elements \eqref{eq:M} for isotopes of a given element with $N$ and $N'$ neutrons, respectively. Moreover, it was pointed out in \cite{fortson_nuclear-structure_1990,pollock_atomic_1992} that isotopic ratios are more sensitive to neutron-distribution uncertainties, compared to measurements in a single isotope. Isotopic ratios are less sensitive to certain types of new physics \cite{ramsey-musolf_low-energy_1999}, on the other hand their reach is complementary to single-isotopes measurements. 
It is thus possible \cite{pollock_atomic_1992, derevianko_reevaluation_2002} to use isotopic comparison to probe neutron skins, test various nuclear-structure models and search for new physics.

This work is organized as follows: in section \ref{sec:ratios} we derive analytical expressions for isotopic ratios \eqref{eq:ratio} in sharp-edge and Fermi distribution models of the nuclear density, and investigate the contributions of neutron skin and new physics to these ratios. Section \ref{sec:deform} is dedicated to a brief discussion of the role of nuclear deformations in spin-independent atomic PNC effects. In section \ref{sec:evaluation} we evaluate the coefficients of different contributions to the nuclear part $\mathcal{N}\tilde{Q}$ in \eqref{eq:M} for a list of atoms of experimental interest for PNC measurements, namely Cs, Ba, Sm, Dy, Yb, Pb, Fr, and Ra. We discuss our results in the section \ref{sec:analysis} and present our conclusions in section \ref{sec:conclusion}.

\section{Isotopic ratios}\label{sec:ratios}
\begin{figure}[b]
	\includegraphics[width=\columnwidth]{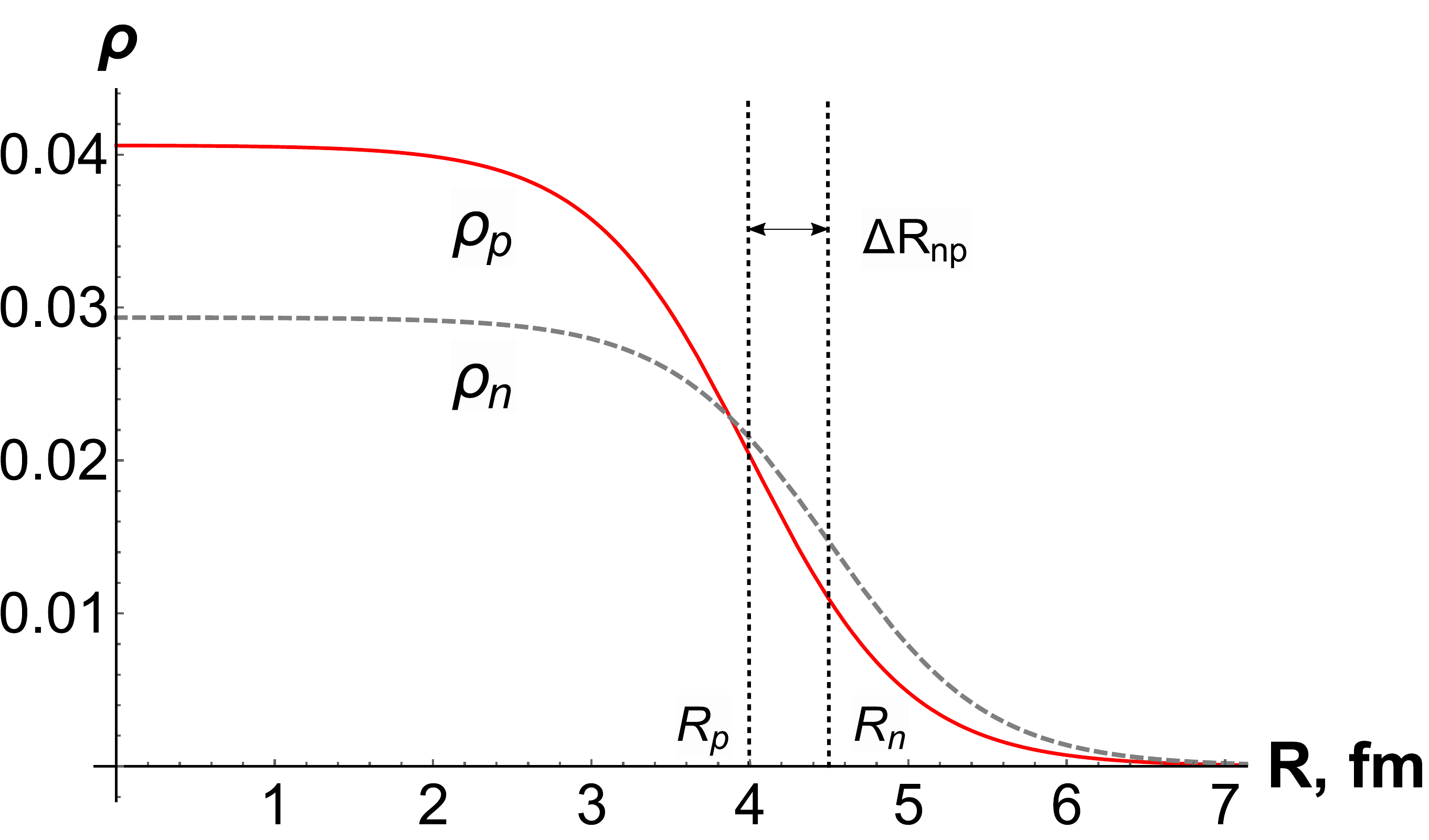}
	\caption{Normalized Fermi distribution of density of nucleons $\rho$, as defined in \eqref{eq:Fermi}, for values of the diffusenes $z=0.5$~fm. Here $R_p$ and $R_n$ are the equivalent radii of proton $\rho_p$ and neutron $\rho_n$ densities and $\Delta R_{np}$ is their difference. 
	}\label{pic:Fermi}
\end{figure}
Let us examine the sensitivity of isotopic ratios to new couplings manifested in the effective nuclear weak charge. Here we split the hypothetical new physics correction $\Delta \tilde{Q}$ to the effective weak charge \eqref{eq:dQ_correction} into proton and neutron contributions:
\begin{equation}
\Delta \tilde{Q} = Z\Delta \tilde{Q}^P+N\Delta \tilde{Q}^N\ .
\end{equation}
For the present purposes, we can roughly assume $\tilde{Q}\approx -N$. To first order of Taylor expansion and denoting $\Delta N=N'-N$, one can write:
\begin{equation}
\frac{\tilde{Q}'}{\tilde{Q}}=\frac{-N'+\Delta \tilde{Q}'}{-N+\Delta \tilde{Q}}\approx\frac{N'}{N}\left[1+\frac{Z\Delta N}{NN'}\Delta \tilde{Q}^P\right] .\label{eq:ratio_newphys}
\end{equation}
Isotopic ratios are therefore primarily sensitive to the new proton couplings $\Delta \tilde{Q}^P$ (or, equivalently, twice as sensitive to the $u$-quark than to the $d$-quark couplings).

Now consider the nuclear-structure corrections \cite{fortson_nuclear-structure_1990,ramsey-musolf_low-energy_1999}. As a first approximation, we use a simplified model of the nucleus: proton and neutron densities are concentric sharp-edge spheres of given radii $R_p$ and $R_n$. 
Using the expansions of Dirac wave functions of $s$ and $p_{1/2}$ electrons inside the nucleus, one can show that the normalized wave-function density $f(r)$ inside the nucleus [see Eq. \eqref{eq:f} and formulas in the Appendix] is:
\begin{equation}
f(r)=1-\frac{1}{2}(Z\alpha)^2\left[\frac{r^2}{R_p^2}-\frac{1}{5}\frac{r^4}{R_p^4}+\frac{1}{75}\frac{r^6}{R_p^6}\right]+\dots
\end{equation}
Performing the integrations in \eqref{eq:q_p} and \eqref{eq:q_n}, one finds:
\begin{equation}
q_p = 1-\frac{817}{3150}(Z\alpha)^2+\mathcal{O}(Z\alpha)^4\ . \label{eq:qp_sphere}
\end{equation}
The function $q_p$ is insensitive to the nuclear parameters $R_p$ and $R_n$ to first order in $(Z\alpha)^2$, and
\begin{multline}
q_n = 1- (Z\alpha)^2\left[\frac{3}{10}\frac{R_n^2}{R_p^2} -\frac{3}{70}\frac{R_n^4}{R_p^4} +\frac{1}{450}\frac{R_n^6}{R_p^6} \right]\\+\mathcal{O}(Z\alpha)^4 .
\end{multline} 
In the sharp-edge spherical model, the radius of the sphere is expressed through the rms radius as [see \eqref{eq:R_SE}]
\begin{equation}
R^2=\frac{5}{3}r^2_\mathrm{rms}\ .
\end{equation}
Let us define a neutron skin parameter using proton $r_p$ and neutron $r_n$ rms radii:
\begin{equation}
y\equiv\frac{r^2_n}{r^2_p}-1\label{eq:y}
\end{equation}
and assume that it is small. We define the thickness of a neutron skin as the difference between neutron and proton rms radii: $\Delta r_\mathrm{np}=r_n-r_p$, then $y\approx 2\Delta r_\mathrm{np}/r_p$. To first order in y \eqref{eq:y} the integral $q_n$ simplifies to:
\begin{align}
q_n 
&= 1-(Z\alpha)^2 \left(\frac{817}{3150}+\frac{116}{525} y \right)+\mathcal{O}(Z\alpha)^4 \nonumber\\
&= q_p-\frac{116}{525} (Z\alpha)^2 y+\mathcal{O}(Z\alpha)^4\ . \label{eq:qn_sphere}
\end{align}
One can evaluate integrals $q_p$ and $q_n$ using a more realistic approach, with the neutron and proton nuclear density having the shape of Fermi distribution:
\begin{equation}
\rho_F(r,R_0)=\mathcal{B}\frac{1}{1+\mathrm{exp}\left(\frac{r-R_0}{z}\right)}\ ,\label{eq:Fermi}
\end{equation}
where $\mathcal{B}$ is normalization constant such that 
\begin{equation}
\int_0^\infty \rho_F(r,R_0)r^2dr=1\ ,
\end{equation}
where the parameter $R_0$ is the radius of the distribution and $z$ the diffuseness (see Fig. \ref{pic:Fermi}). Here the rms radius is expressed through $R_0$ as:
\begin{equation}
r^2_\mathrm{rms}=\langle r^2\rangle\approx \frac{3}{5}R_0^2+\frac{7}{5}\pi^2 z^2\ .
\end{equation}
In the limit of $z\rightarrow 0$ the $\rho_F(R,R_0)$ reduces to the sharp-edge spherical distribution with radius $R_0$. For further calculations, we use the mean value of diffuseness of experimental charge distributions $z= 0.53(4)$~fm presented in Ref. \cite{chamon_toward_2002}. Using the approach similar to that employed for obtaining Eqs.~\eqref{eq:qp_sphere} and \eqref{eq:qn_sphere}, one can show that to first order in $(Z\alpha)^2$, integrals \eqref{eq:q_p} and \eqref{eq:q_n} can be expressed as:
\begin{align}
q_p&=1-\varepsilon(Z\alpha)^2+\mathcal{O}(Z\alpha)^4\ \label{eq:qp_Fermi} ,\\ 
q_n&=q_p-\eta y(Z\alpha)^2+\mathcal{O}(Z\alpha)^4\ ,\label{eq:qn_Fermi}
\end{align}
assuming $|y|\ll 1$. Here $\varepsilon$ and $\eta$ become functions of the nuclear radius $R$, but remain close to the sharp-edge values $\varepsilon_0=817/3150\approx 0.2594$ and $\eta_0=116/525\approx 0.2210$ (see Table \ref{tab:Fermi_coef}). In fact, the dependence of $\varepsilon$ and $\eta$ on $R$ is weak and they can be treated as constants for a selected element. A more detailed approach to analytic calculation of the dependence of a PNC matrix element on the nuclear shape was described in \cite{james_parametric_1999}.

\begin{table}[tb]
\caption{Mean values of the coefficients $\overline{\varepsilon}$ from \eqref{eq:qp_Fermi} and $\overline{\eta}$ from \eqref{eq:qn_Fermi} for different elements, assuming Fermi distribution of nuclear density with the diffuseness value $z=0.53(4)$~fm \cite{chamon_toward_2002}. Here $\Delta\varepsilon$ and $\Delta\eta$ represent the maximum deviation of $\varepsilon$ and $\eta$ from the mean value in isotopes $A$ of a given element. For comparison, in the sharp-edge model, $\varepsilon_0\approx 0.2594$ and $\eta_0\approx 0.2210$. Thus, generally, $\varepsilon>\varepsilon_0$ and $\eta>\eta_0$. }\label{tab:Fermi_coef}
\begin{tabular}{p{0.5cm} p{0.5cm} p{1.6cm} p{1.2cm} p{1.3cm} p{1.2cm} p{1.1cm}}
 	&	$Z$	&	$A$	&   $\overline{\varepsilon}$	&	$\Delta\varepsilon$	&	$\overline{\eta}$	&	$\Delta\eta$	\\
\hline
Cs  &   55  &   $131-137$   &   0.2974  &   0.0001  &   0.24487 &   0.00007 \\
Ba	&	56	&   $130-138$	&	0.2970	&   0.0001	&	0.24460	&	0.00007	\\
Sm	&	62	&	$144-154$	&	0.294	&	0.001	&	0.2427	&   0.0007	\\
Dy	&	66	&	$156-164$	&	0.2913	&	0.0005	&	0.2412	&	0.0003	\\
Yb	&	70	&	$168-176$	&	0.2900	&	0.0004	&	0.2404	&	0.0002	\\
Pb	&	82	&	$202-208$	&	0.2877	&	0.0002	&	0.2390	&	0.0001	\\
Fr	&	87	&	$207-228$	&	0.2859	&	0.0009	&	0.2379	&	0.0006	\\
Ra	&	88	&	$208-232$	&	0.286	&	0.001	&	0.2378	&	0.0006	\\
\hline\hline										
\end{tabular}
\end{table}

Let us estimate the influence of nuclear parameters $R_p$ and $R_n$ on the isotopic ratio. For this analysis we assume no new physics couplings ($\Delta\tilde{Q}=0$). With $(1-4\sin^2\theta)\equiv \xi $ the ratio \eqref{eq:ratio} reads:
\begin{align}
\mathcal{R}&=\left(\frac{R'_p}{R_p}\right)^{2\gamma-2}\frac{Z \xi q'_p-N' q'_n}{Z \xi q_p-N q_n}\nonumber\\ &=\left(\frac{R'_p}{R_p}\right)^{2\gamma-2} \frac{N'}{N}\varkappa\ .
\end{align}
Here we retained only the first order in the expansion of $\mathcal{R}$ in $(Z\alpha)^2$ and introduced the notations: $\Delta N=N'-N$ and $\varkappa$:
\begin{equation}
\varkappa\equiv \frac{q'_p}{q_p}\left[1+\frac{Z\xi\Delta N}{N N'}-(Z\alpha)^2 \left(\frac{\eta' y'}{q'_p}-\frac{\eta y}{q_p}\right)\right]\ .\label{eq:ratio_nsk}
\end{equation}

If one assumes the sharp-edge model where $q'_p=q_p=q_{p,0}$ and $\eta'=\eta=\eta_0$, then $\varkappa$ simplifies to:
\begin{equation}
\varkappa_0=1+\frac{Z\xi\Delta N}{N N'}-\frac{\eta_0}{q_{p,0}} (Z\alpha)^2 (y'-y)\ .\label{eq:ratio_nsk_SE}
\end{equation}
Here the second term comes from the proton contribution to the nuclear effective weak charge and it does not depend on the nuclear structure. The third term gives the explicit dependence of $\kappa_0$ on the change of the ratio of neutron and proton radii; in the sharp-edge model, if the neutron radius is strictly proportional to proton radius, $y'-y$ sums to zero.

Let us write an expansion of the isotopic ratio \eqref{eq:ratio}, which combines both new physics \eqref{eq:ratio_newphys} and neutron skin \eqref{eq:ratio_nsk} contributions, in order to examine their joined contributions to the ratio ($N'>N$):
\begin{gather}
\mathcal{R}\approx\left( \frac{R'_p}{R_p}\right)^{2\gamma-2} \frac{\tilde{Q}'}{\tilde{Q}}=
\left(\frac{R'_p}{R_p}\right)^{2\gamma-2}\frac{N'}{N}\frac{q'_p}{q_p}\times\nonumber\\
\Big[ 1+\frac{Z\xi\Delta N}{NN'}-(Z\alpha)^2\left(\frac{\eta' y'}{q'_p}-\frac{\eta y}{q_p}\right)+ \nonumber\\
 \Delta \tilde{Q}^N\frac{q'_p-q_p}{q'_p q_p}+Z\Delta \tilde{Q}^P\frac{N'q_p'-Nq_p}{N'Nq'_p q_p} \Big]\ .\label{eq:ratio_Fermi_general}
\end{gather}
In the sharp-edge model:
\begin{gather}
\mathcal{R}\approx
\left(\frac{R'_p}{R_p}\right)^{2\gamma-2}\frac{N'}{N}\times\nonumber\\
\left[1+\frac{Z\Delta N}{NN'}\left(\xi+\frac{\Delta \tilde{Q}^P}{q_{p,0}}\right)-\frac{\eta_0}{q_{p,0}}(Z\alpha)^2 (y'-y)\right]\ .\label{eq:ratio_SE}
 \end{gather}
As Table \ref{tab:Fermi_coef} indicates, $\varepsilon$ and $\eta$ are almost constant for a given element, and therefore the values of $q_p$, $q'_p$ and $\eta$, $\eta'$ are close. The expressions \eqref{eq:ratio_nsk_SE} and \eqref{eq:ratio_SE}, initially derived for sharp-edge model, can thus also be used within the Fermi distribution model, provided that $q_{p,0}$ is replaced with $\overline{q_p}$ and $\eta_0$ with $\overline{\eta}$ (see Table \ref{tab:Fermi_coef}). The Eq.~\eqref{eq:ratio_SE} then becomes
\begin{gather}
\mathcal{R}\approx
\left(\frac{R'_p}{R_p}\right)^{2\gamma-2}\frac{N'}{N}\times\nonumber\\
\left[1+\frac{Z\Delta N}{NN'}\left(\xi+\frac{\Delta \tilde{Q}^P}{\overline{q_p}}\right)-\frac{\overline{\eta}}{\overline{q_p}}(Z\alpha)^2 (y'-y)\right]\ .\label{eq:ratio_SE_withFermi}
\end{gather}

Note that $\Delta\tilde{Q}^P$ enters the isotopic ratio \eqref{eq:ratio_SE_withFermi} the same way as $\xi$. New physics corrections to proton coupling would thus manifest themselves similarly to corrections to Weinberg angle. In fact, probing the Weinberg angle in low-energy experiments presents an unique opportunity for search for a number of new physics scenarios, such as dark Z boson \cite{davoudiasl_muon_2014,safronova_search_2018}.

The characteristic scales of $\Delta \tilde{Q}^P$ for various new physics scenarios were discussed e.g. in \cite{ramsey-musolf_low-energy_1999}; for a more recent review see \cite{safronova_search_2018}. Constraints on $\Delta Q^P$ arising in presence of additional $Z'$ bosons, were recently reported in \cite{antypas_isotopic_2019}, and were based on measurements of Yb PNC amplitudes and atomic calculations \cite{dzuba_probing_2017}.

\section{Nuclear deformation}\label{sec:deform}
As Eq.\,\eqref{eq:ratio_SE_withFermi} shows, the value of the ratio $\mathcal{R}$ depends primarily on the neutron numbers and radii of the involved isotopes. It is thus most efficient to make use of a pair of isotopes with the largest possible $\Delta N$ to maximize the ratio. 

A large change in rms radius between two isotopes can occur when there is a significant change in nuclear deformation. The simplest case of quadrupole deformation of a sharp-edge nucleus is a surface:
\begin{equation}
r=r_0\left(1+\beta Y^0_2\right)\ ,\label{eq:deform}
\end{equation}
where $r_0$ is a radius parameter and $Y^0_2$ a spherical harmonic
\begin{equation}
Y^0_2=\sqrt{\frac{5}{16\pi}}\left(3 \cos^2\theta-1\right).
\end{equation}
Here $\beta$ is the deformation parameter; usually, $\beta\lesssim 1/3$ \cite{janecke_simple_1981}. The mean charge radius that corresponds to the surface \eqref{eq:deform} would be expressed as:
\begin{equation}
    \left<r^2\right>_\beta=\frac{3}{5}r_0^2\left(1+\frac{7}{4\pi}\beta^2\right)\ .
\end{equation}
Given the constant volume of the nucleus, the mean charge radii of the deformed $\langle r^2\rangle_\beta$ and of the initial spherical nucleus $\langle r^2\rangle_0$ are related through:
\begin{equation}
\left<r^2\right>_\beta=\left<r^2\right>_0\left(1+\frac{5}{4\pi}\beta^2\right)\ .
\end{equation}
The quadrupole deformation is directly connected to the nuclear intrinsic quadrupole moment $Q$ \cite{king_isotope_2013}:
\begin{equation}
Q=\frac{3}{\sqrt{5\pi}}Zr^2_0\beta(1+0.36\beta)\ .
\end{equation}
There could be a weak quadrupole moment as well, due to the deviation of neutron distribution from the spherical shape. It would result in a tensor PNC interaction between the nucleus and the electrons, as discussed in Refs. \cite{sushkov_o._1978,flambaum_effect_2017}. A tensor interaction has different selection rules from a scalar (spin-independent) interaction, so it is possible to separate one from another in an experiment. Therefore, for the purposes of the present work, it is sufficient to treat deformed nuclei as spherical nuclei with equivalent charge radius $R^2_p$.

\section{Evaluation of parameters in PNC matrix element for specific isotopes}\label{sec:evaluation}
\begin{figure}[bt]
	\includegraphics[width=\columnwidth]{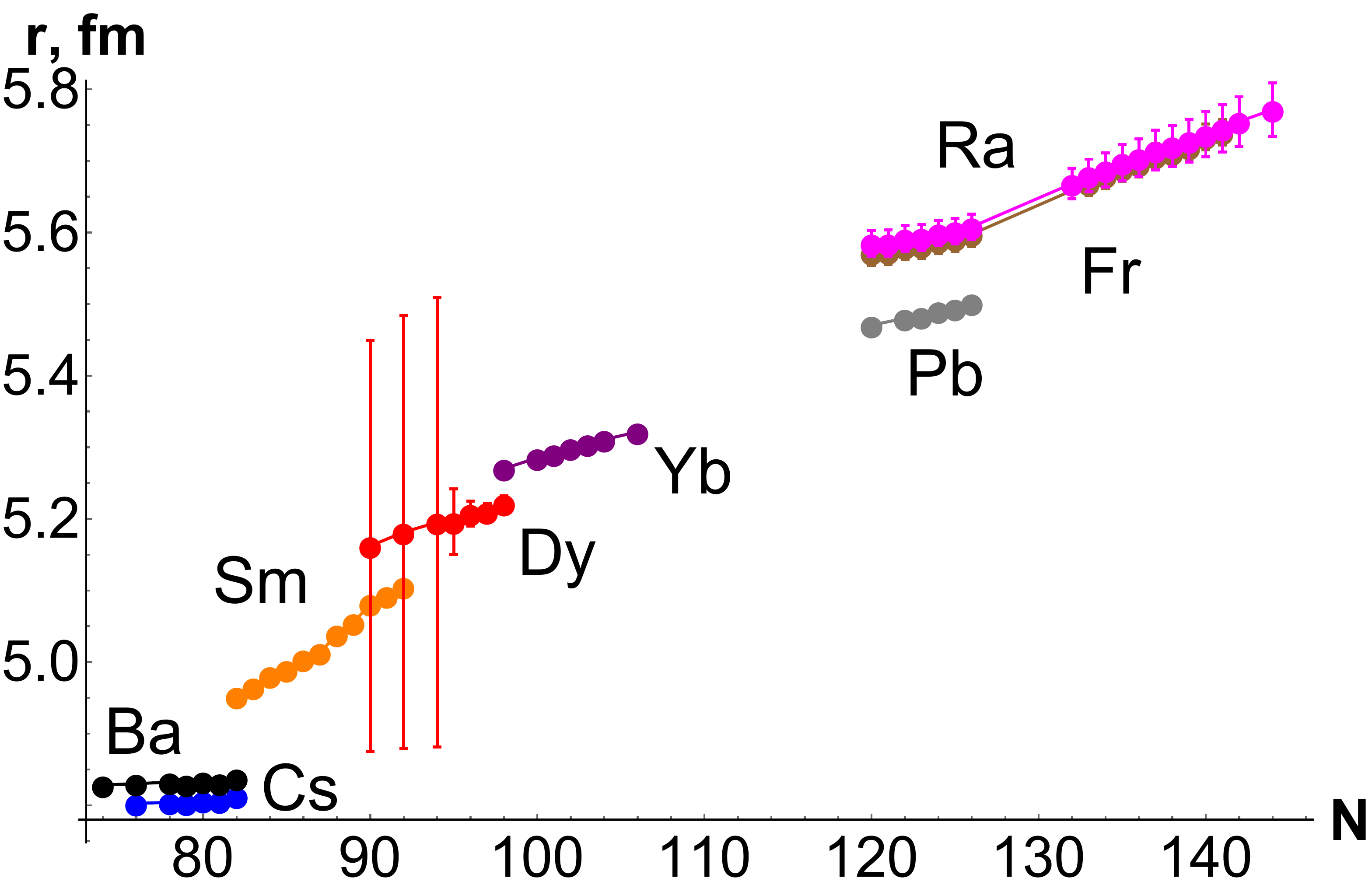}
	\caption{Experimental rms proton radii $(r)$ for eight elements, as listed in \cite{angeli_table_2013} and Table \ref{tab:results}, plotted against the neutron numbers $(N)$ of each element. On the scale of this plot, uncertainties are visible only for Dy and Ra. The points for Ra and Fr almost completely overlap. The first isotopes in the lists of elements are: $^{131}$Cs, $^{130}$Ba, $^{144}$Sm, $^{156}$Dy, $^{168}$Yb, $^{202}$Pb, $^{207}$Ra and $^{208}$Fr. It appears that the uncertainties listed in \cite{angeli_table_2013} may be overestimated and there may be correlations in the radii of different isotopes of a given element.} \label{pic:radii}
\end{figure}
Let us find the dependence of the nuclear part in \eqref{eq:M} on the ratio of the neutron and proton equivalent radii $R^2_n/R^2_p$ and on the new-physics couplings to protons and neutrons in isotopes proposed for PNC experiments \cite{safronova_search_2018}. 
The neutron-skin parameter $y$ is defined in Eq.\,\eqref{eq:y}. To first order in $(Z\alpha)^2$ and $y$, the nuclear part of \eqref{eq:M} can be expressed as:
\begin{align}
\mathcal{N}\tilde{Q}& \nonumber\\
=&\left(\frac{2 Z R_p}{a_B}\right)^{2\gamma -2}\Big[ Z q_p(1-4\sin^2\theta)-N q_p \nonumber \\
+&\eta N(Z\alpha)^2y+Z \Delta \tilde{Q}^P+N \Delta \tilde{Q}^N \Big] \nonumber\\
\equiv & \left(\frac{2 Z R_p}{a_B}\right)^{2\gamma -2}\left[B_0+B_1 y+Z \Delta \tilde{Q}^P+N \Delta \tilde{Q}^N\right]\ ,\label{eq:nuclear}
\end{align}
defining
\begin{align}
B_0&=\left[Z(1-4\sin^2\theta)-N\right]q_p\ ,  \label{eq:b0} \\
B_1&=\eta N (Z\alpha)^2\ . \label{eq:b1}
\end{align}
For our calculations, we use the Fermi distribution of nucleons. We substitute the corresponding values of $N$, $Z$, $r_p$ and $z=0.53(4)$~fm for isotopes of Cs, Ba, Sm, Dy, Yb, Pb, Fr and Ra into Eq. (\ref{eq:nuclear}), compute the coefficients $B_0$ and $B_1$ and list the results in Table \ref{tab:results}. The rms nuclear charge radii $r_p$ are taken from Ref. \cite{angeli_table_2013}. The errors presented in Table \ref{tab:results} come from the predominantly experimental errors in $r_p$ and the error in the value of $z$.  
If one used SE model in contrast to the Fermi distribution described above, the normalization $\mathcal{N}=(2 Z R_p/a_B)^{2\gamma-2}$ would be lower by $2$-$3\%$, the values of $q_p$ and $B_0$ would change by approximately $1\%$, whereas the coefficient $B_1$ would decrease by $7-10\%$. 

Let us examine the role of experimental uncertainties in nuclear charge radius in a possible extraction of neutron-skin thickness. Based on neutron-skin calculations \cite{brown_calculations_2009}, a typical value of the parameter $y$ for the isotopes considered in the present work would be $0.04-0.09$. If one assumes that both the nuclear part of the PNC amplitude $\mathcal{N}\tilde{Q}$ and the weak charge $Q_W$ are known with a $0.1\%$ certainty, whereas the experimental uncertainties of charge radii are as listed in \cite{angeli_table_2013} and $z=0.53(4)$, then the error in $y$ is $\Delta y\approx 0.03-0.05$, with the tendency of $\Delta y$ to decrease with $Z$. This error is typically more than $50\%$ of the value of $y$, notably for lighter atoms. If the uncertainty of $\mathcal{N}\tilde{Q}$ is raised to $0.3\%$, the error in $y$ becomes more than its value in lighter atoms and more than a half of its value in heavy atoms. It is thus crucial for the extraction of $y$ from a single-isotope amplitude $\mathcal{N}\tilde{Q}$ to know with high accuracy both the charge radius and the diffuseness of the nuclei. Indeed, a large portion of the resulting error comes from the uncertainty in $z$. If one sets $\Delta z=0$ (given the case of $0.1\%$ certainty for $\mathcal{N}\tilde{Q}$ and $Q_W$), the reduction in $\Delta y$ is $30-50\%$. Interpreting experimental results thus poses a challenge for nuclear models.

\section{Analysis and outlook}\label{sec:analysis}
\subsection{Single isotope}
It is convenient to express the size of nuclear-structure corrections as a fraction of the dominant PNC effect arising solely from $\hat{h}_{SI,0}$ [Eq.\,\eqref{eq:h_SI_0}]:
\begin{gather}
\mathcal{M}_0=\langle j|\hat{h}_{SI,0}|i\rangle=\frac{G_f}{2\sqrt{2}}(2C_{1p}Z +2C_{1n} N) I_p\ ,\\
\mathcal{M}_1=\langle j|\hat{h}_{SI}|i\rangle=\frac{G_f}{2\sqrt{2}}(2C_{1p}Z I_p +2C_{1n} N I_n)\ ,\\
I_p=\int\rho_p(r)\psi^\dagger_j\gamma_5\psi_i d^3r\ ,\\
I_n=\int\rho_n(r)\psi^\dagger_j\gamma_5\psi_i d^3r\ ,
\end{gather}
thus leading to
\begin{equation}
\frac{\delta E^\mathrm{n.s.}_\mathrm{PNC}}{E_\mathrm{PNC}}=\frac{\mathcal{M}_1-\mathcal{M}_0}{\mathcal{M}_0}=\frac{2 C_{1n}N}{Q_W}\left(\frac{I_n}{I_p}-1\right)\ ,\label{eq:dEE_general}
\end{equation}
where $Q_W$ is the SM nuclear weak charge defined in Eq.\,\eqref{eq:Q_W_traditional}. Assuming, as most realistic nuclear models imply, that the distributions of protons and neutrons are not radically different in shape, we express first-order values of $q_p$ and $q_n$ in the form of \eqref{eq:qp_Fermi} and \eqref{eq:qn_Fermi}.  
At the tree level and using the electron wave functions given in the Appendix, we obtain
\begin{align}
\frac{\delta E^\mathrm{n.s.}_\mathrm{PNC}}{E_\mathrm{PNC}}&=\frac{-N}{Q_W}\left(\frac{q_n}{q_p}-1\right)=\frac{N\eta y}{Q_W q_p}(Z\alpha)^2\nonumber \\&=\frac{B_1}{Q_W q_p}y\equiv Ky\ .\label{eq:K}
\end{align}
Since $2 K/(Z\alpha)^2 \approx 3/7$, the quantity \eqref{eq:K} is approximately equal to that given by the commonly used formula (see e.g.~\cite{brown_calculations_2009}): 
\begin{equation}
\frac{\delta E^\mathrm{n.s.}_\mathrm{PNC}}{E_\mathrm{PNC}}=-\frac{3}{7}(Z\alpha)^2\frac{\Delta R_{np}}{R_p}\ ,\label{eq:3/7}
\end{equation}
which comes from a more crude estimate of integrals $q_p$ \eqref{eq:q_p} and $q_n$ \eqref{eq:q_n} in the sharp-edge model. Eq.\,\eqref{eq:3/7} slightly underestimates the effect of a neutron skin on the PNC amplitude $E_\mathrm{PNC}$.

We estimate the ratio $\delta E^\mathrm{n.s.}_\mathrm{PNC}/E_\mathrm{PNC}$ according to Eq.~\eqref{eq:dEE_general}, using neutron skin thicknesses calculated in \cite{brown_calculations_2009} and tree-level $Q_W$ \eqref{eq:QW_tree}, taking into account uncertainties of $r_0$ and $z$. The results are presented in third-to-last column of Table \ref{tab:results}. The values of $\delta E^\mathrm{n.s.}_\mathrm{PNC}/E_\mathrm{PNC}$ calculated with the simplified formula \eqref{eq:K} would differ only insignificantly.

\subsection{Isotopic ratio}
From Eqs.\,\eqref{eq:ratio_Fermi_general} and \eqref{eq:ratio_SE} one can see that isotopic ratios are, indeed, largely insensitive to the new couplings to neutron $\Delta \tilde{Q}^N$. This allows one to isolate new proton couplings when taking a ratio of PNC effects in different isotopes. 

The relative sensitivity to nuclear structure and new proton couplings in isotopic ratios is most obvious if we turn to the formula \eqref{eq:ratio_SE_withFermi} and compare the coefficients
\begin{gather}
c_p\equiv \frac{Z\Delta N}{NN'}\ , \label{eq:cp}\\
c_n\equiv -\frac{\overline{\eta}}{\overline{q_p}}(Z\alpha)^2\ .\label{eq:cn}
\end{gather}
Both coefficients \eqref{eq:cp} and \eqref{eq:cn} are, generally, of the order of $0.01$. 
In lighter atoms, such as Cs and Ba, $c_p$ can be greater than $c_n$, which implies slightly better sensitivity to new proton couplings, but in heavier atoms (Dy, Yb, Pb, Fr and Ra) $c_n$ dominates. 

To give an estimate for the relative size of proton and neutron contribution in the isotopic ratio, we substitute $(y'-y)$ based on the neutron skin calculations \cite{brown_calculations_2009} and compare second and third terms in parentheses in Eq.~\eqref{eq:ratio_SE_withFermi}. The ratio of neutron to proton contribution to \eqref{eq:ratio_SE_withFermi} is
\begin{equation}
\chi=\frac{c_n(y'-y)}{c_p\xi}\label{eq:chi}
\end{equation}
is $\approx 0.2$ for all isotopes of Cs and Ba, $\approx 0.3$ for Sm and Dy and varies between $0.3$ and $0.6$ for selected pairs of isotopes in Yb, Fr and Ra. The neutron distribution effect is thus more pronounced in heavier atoms. If one aims to avoid significant neutron skin contributions, it is better to choose lighter elements.

Since the equation \eqref{eq:ratio_SE_withFermi} provides a good approximation for the Fermi model and the neutron-skin correction to the ratio depends solely on 
\begin{equation}
y'-y=\left(\frac{R'_n}{R'_p}\right)^2-\left(\frac{R_n}{R_p}\right)^2\ ,
\end{equation}
it means that the errors in neutron skins for different isotopes, if correlated, cancel out when an isotopic ratio is evaluated. The errors in $y$ are indeed shown to be correlated in Ref. \citep{brown_calculations_2009}. This cancellation enables use of isotopic ratio measurements to search for new physics couplings.

If the aim is to detect the neutron skin effect in an isotopic ratio, it is meaningful to choose pairs of isotopes with greater relative neutron skin contribution:
\begin{equation}
\frac{\Delta_\mathrm{n.s.}}{\mathcal{R}}=\frac{c_n (y'-y)}{1+c_p\xi+c_n(y'-y)}\ .\label{eq:delta_nsk}
\end{equation}
Values of the above ratio for selected pairs of isotopes are presented in Table \ref{tab:results}.

A $0.1\%$ uncertainty of nuclear weak charge $Q_W$ would result in approximately $0.14\%$ uncertainty of the ratio $\mathcal{R}$ for any pair of isotopes considered, which is larger than or comparable to the neutron skin effect \eqref{eq:delta_nsk} in some of the pairs, but is about a half of $\Delta_\mathrm{n.s.}/\mathcal{R}$ in others, prominently in heavier elements.

Note that the parameter $c_n$ [Eq.\,\eqref{eq:cn}] is close in value to the sensitivity to neutron skin $K$ [Eq.\,\eqref{eq:K}] of an individual isotope, as indeed follows from their definition. Therefore in Table \ref{tab:results} we present only $c_p$ and $c_n$.

\LTcapwidth=\textwidth
\begin{longtable*}[b]{p{1.2cm} p{0.7cm} p{0.5cm} p{1.4cm} p{1.4cm} p{1.3cm} p{1.5cm}  p{1.2cm} p{0.6cm} p{1.0cm} | p{1.4cm} p{1.6cm} p{1.3cm} p{0.5cm}}
\\[0.7cm]
\caption{Evaluated coefficients from Eqs.\,\eqref{eq:nuclear}--\eqref{eq:b1} for the nuclear factor of the PNC amplitude, expanded in Eq.\,\eqref{eq:nuclear}. The parameter $Z$ is the atomic number of an element, $A$ is the atomic mass number, $N$ the number of neutrons and $r_p$ is the rms radius of a given isotope in fm \cite{angeli_table_2013}. The normalization constant is $\mathcal{N}=(2 Z R_p/a_B)^{2\gamma-2}$, where $R_p$ is the equivalent proton radius. The coefficients $B_0$ and $B_1$ are defined in Eqs.\,\eqref{eq:b0} and \eqref{eq:b1}. The isotopic-ratio parameters $c_p$ and $c_n$ follow Eqs.\,\eqref{eq:cp} and \eqref{eq:cn}. The initial reference isotope for $c_p$ is always the first in the list belonging to each element, and $c_n$ is the same for any pair of isotopes of a given element. 
Calculated differences between neutron and proton rms radii $\Delta r_{np}$ come from Ref. \cite{brown_calculations_2009}. Approximate values of neutron-skin contributions to single-isotope PNC amplitudes $\delta E^\mathrm{n.s.}/E$ [Eq.\,\eqref{eq:K}] and to the ratios of PNC amplitudes $\Delta_\mathrm{n.s.}/\mathcal{R}$ [Eq.\,\eqref{eq:delta_nsk}] are listed in the third-to-last and second-to-last columns. The last column indicates ratios of neutron skin contributions to proton contributions $\chi$ in an isotopic ratio, defined in \eqref{eq:chi}, for a given pair of isotopes. The reference isotope of an element for $\Delta_\mathrm{n.s.}/\mathcal{R}$ and $\chi$ is always the first isotope with known $\Delta r_{np}$. Stable isotopes and isotopes with half-life longer than $10^8$~y are written in boldface. }\label{tab:results} \\

\hline
Element	&	$A$	&	$N$	&	$r_p$, fm	&$\mathcal{N}$	&	$q_p$		&	$B_0$		&	$B_1$	& $c_p$ 	&	$c_n$	& $\Delta r_{np}$, fm & $\delta E^\mathrm{n.s.}/E$ & $\Delta_\mathrm{n.s.}/\mathcal{R}$	& $\chi$\\
\hline                                                                                    
                    
Cs		&	$131$			&	$76$			&	$4.803(5)$	&	$2.111(5)$	&	$0.952(1)$	&	$-68.17(7)$	&	$	3.00(5)$	&	$-	$	&	$-0.04$	&	$0.139(34)$&	$-0.0026(6)	$	&	$	-	$	& $-$ \\		
$Z=55$	&	$\mathbf{133}$	&	$\mathbf{78}$	&	$4.804(5)$	&	$2.111(5)$	&	$0.952(1)$	&	$-70.07(8)$	&	$	3.08(5)$	&	$0.02$	&			&	$0.158(37)$&	$-0.0029(7)	$	&	$-0.0003$ & $0.2$	\\					
		&	$134$			&	$79$			&	$4.803(5)$	&	$2.111(5)$	&	$0.952(1)$	&	$-71.03(8)$	&	$	3.12(5)$	&	$0.03$	&			&	\\											
		&	$135$			&	$80$			&	$4.807(5)$	&	$2.111(5)$	&	$0.952(1)$	&	$-71.98(8)$	&	$	3.16(5)$	&	$0.04$	&			&	$0.176(40)$&	$-0.0032(7)	$	&	$-0.0007$ & $0.2$	\\					
		&	$136$			&	$81$			&	$4.806(5)$	&	$2.111(5)$	&	$0.952(1)$	&	$-72.93(8)$	&	$	3.20(5)$	&	$0.04$	&			&	\\											
		&	$137$			&	$82$			&	$4.813(5)$	&	$2.110(5)$	&	$0.952(1)$	&	$-73.88(8)$	&	$	3.23(5)$	&	$0.05$	&			&	$0.193(42)$&	$-0.0035(8)	$	&	$-0.0010$ & $0.2$	\\					
\hline																																
Ba		&	$\mathbf{130}$	&	$\mathbf{74}$	&	$4.828(5)$	&	$2.163(6)$	&	$0.950(1)$	&	$-66.07(7)$	&	$	3.02(5)	$	&	$-	$	&	$-0.04$	&	$0.104(29)$&	$-0.0020(6)	$	&	$	-	$   & $-$	\\		
$Z=56$	&	$\mathbf{132}$	&	$\mathbf{76}$	&	$4.830(5)$	&	$2.163(6)$	&	$0.950(1)$	&	$-67.97(8)$	&	$	3.10(5)	$	&	$0.02$	&			&	$0.124(32)$&	$-0.0024(6)	$	&	$-0.0004$    & $0.2$	\\					
		&	$133$			&	$77$			&	$4.829(5)$	&	$2.163(6)$	&	$0.950(1)$	&	$-68.92(8)$	&	$	3.15(5)	$	&	$0.03$	&			&	\\											
		&	$\mathbf{134}$	&	$\mathbf{78}$	&	$4.832(5)$	&	$2.163(6)$	&	$0.950(1)$	&	$-69.87(8)$	&	$	3.19(5)	$	&	$0.04$	&			&	$0.143(34)$&	$-0.0027(6)	$	&	$-0.0007$    & $0.2$	\\					
		&	$\mathbf{135}$	&	$\mathbf{79}$	&	$4.829(5)$	&	$2.163(6)$	&	$0.950(1)$	&	$-70.82(8)$	&	$	3.23(5)	$	&	$0.05$	&			&	\\											
		&	$\mathbf{136}$	&	$\mathbf{80}$	&	$4.833(5)$	&	$2.163(6)$	&	$0.950(1)$	&	$-71.78(8)$	&	$	3.27(5)	$	&	$0.06$	&			&	$0.161(37)$&	$-0.0030(7)	$	&	$-0.0010$    & $0.2$	\\					
		&	$\mathbf{137}$	&	$\mathbf{81}$	&	$4.831(5)$	&	$2.163(6)$	&	$0.950(1)$	&	$-72.73(8)$	&	$	3.31(5)	$	&	$0.07$	&			&	\\											
		&	$\mathbf{138}$	&	$\mathbf{82}$	&	$4.838(5)$	&	$2.162(6)$	&	$0.950(1)$	&	$-73.68(8)$	&	$	3.35(5)	$	&	$0.07$	&			&	$0.179(40)$&	$-0.0034(8)	$	&	$-0.0014$    & $0.2$	\\					
\hline																																
Sm		&	$\mathbf{144}$	&	$\mathbf{82}$	&	$4.952(3)$	&	$2.529(8)$	&	$0.940(1)$	&	$-72.39(9)$	&	$	4.09(6)	$	&	$-	$	&	$-0.05$	&	$0.098(27)$&	$-0.0022(6)	$	&	$	-	$   & $-$	\\		
$Z=62$	&	$145$			&	$83$			&	$4.965(3)$	&	$2.527(8)$	&	$0.940(1)$	&	$-73.3(1)$	&	$	4.13(6)	$	&	$0.01$	&			&	\\										
		&	$146$			&	$84$			&	$4.981(4)$	&	$2.525(8)$	&	$0.940(1)$	&	$-74.3(1)$	&	$	4.18(6)	$	&	$0.02$	&			&	$0.122(32)$&	$-0.0028(7)	$	&	$-0.0005$    & $0.4$	\\				
		&	$\mathbf{147}$	&	$\mathbf{85}$	&	$4.989(4)$	&	$2.524(8)$	&	$0.940(1)$	&	$-75.2(1)$	&	$	4.23(6)	$	&	$0.03$	&			&	\\										
		&	$\mathbf{148}$	&	$\mathbf{86}$	&	$5.004(3)$	&	$2.522(8)$	&	$0.940(1)$	&	$-76.2(1)$	&	$	4.28(6)	$	&	$0.04$	&			&	$0.144(35)$&	$-0.0032(8)	$	&	$-0.0010$    & $0.3$	\\				
		&	$\mathbf{149}$	&	$\mathbf{87}$	&	$5.013(4)$	&	$2.521(8)$	&	$0.940(1)$	&	$-77.1(1)$	&	$	4.32(6)	$	&	$0.04$	&			&	\\										
		&	$\mathbf{150}$	&	$\mathbf{88}$	&	$5.039(5)$	&	$2.518(7)$	&	$0.940(1)$	&	$-78.0(1)$	&	$	4.37(6)	$	&	$0.05$	&			&	$0.166(39)$&	$-0.0037(9)	$	&	$-0.0014$    & $0.3$	\\				
		&	$151$			&	$89$			&	$5.055(6)$	&	$2.516(7)$	&	$0.940(1)$	&	$-79.0(1)$	&	$	4.42(6)	$	&	$0.06$	&			&	\\										
		&	$\mathbf{152}$	&	$\mathbf{90}$	&	$5.082(6)$	&	$2.512(7)$	&	$0.940(1)$	&	$-79.9(1)$	&	$	4.46(6)	$	&	$0.07$	&			&	$0.187(43)$&	$-0.004(1)	$	&	$-0.0018$    & $0.3$	\\				
		&	$153$			&	$91$			&	$5.093(7)$	&	$2.511(7)$	&	$0.940(1)$	&	$-80.9(1)$	&	$	4.51(6)	$	&	$0.07$	&			&	\\										
		&	$\mathbf{154}$	&	$\mathbf{92}$	&	$5.105(7)$	&	$2.509(7)$	&	$0.940(1)$	&	$-81.8(1)$	&	$	4.56(6)	$	&	$0.08$	&			&	$0.219(48)$&	$-0.005(1)	$	&	$-0.0025$    & $0.4$	\\				
\hline																																
Dy		&	$\mathbf{156}$	&	$\mathbf{90}$	&	$5.2(3)	$	&	$2.81(5)$	&	$0.932(2)$	&	$-79.0(1)$	&	$	5.04(9)	$	&	$-	$	&	$-0.06$	&	$0.135(35)$&	$-0.0033(9)	$	&	$	-	$   & $-$	\\
$Z=66$	&	$\mathbf{158}$	&	$\mathbf{92}$	&	$5.2(3)	$	&	$2.80(5)$	&	$0.932(2)$	&	$-80.9(1)$	&	$	5.15(9)	$	&	$0.02$	&			&	$0.150(37)$&	$-0.004(1)	$	&	$-0.0003$    & $0.3$	\\			
		&	$\mathbf{160}$	&	$\mathbf{94}$	&	$5.2(3)	$	&	$2.80(5)$	&	$0.932(2)$	&	$-82.7(1)$	&	$	5.26(9)	$	&	$0.03$	&			&	$0.164(38)$&	$-0.004(1)	$	&	$-0.0007$    & $0.3$	\\			
		&	$\mathbf{161}$	&	$\mathbf{95}$	&	$5.20(5)$	&	$2.80(1)$	&	$0.932(1)$	&	$-83.7(1)$	&	$	5.31(7)	$	&	$0.04$	&			&	\\										
		&	$\mathbf{162}$	&	$\mathbf{96}$	&	$5.21(2)$	&	$2.801(9)$	&	$0.932(1)$	&	$-84.6(1)$	&	$	5.37(7)	$	&	$0.05$	&			&	$0.178(40)$&	$-0.004(1)	$	&	$-0.0010$    & $0.3$	\\				
		&	$\mathbf{163}$	&	$\mathbf{97}$	&	$5.21(1)$	&	$2.800(9)$	&	$0.932(1)$	&	$-85.5(1)$	&	$	5.42(7)	$	&	$0.05$	&			&	\\										
		&	$\mathbf{164}$	&	$\mathbf{98}$	&	$5.22(1)$	&	$2.799(9)$	&	$0.933(1)$	&	$-86.5(1)$	&	$	5.48(7)	$	&	$0.06$	&			&	$0.191(42)$&	$-0.005(1)	$	&	$-0.0013$    & $0.3$	\\				
\hline																																	
Yb		&	$\mathbf{168}$	&	$\mathbf{98}$	&	$5.270(6)$	&	$3.15(1)$	&	$0.924(1)$	&	$-85.4(1)$	&	$	6.15(8)	$	&	$-	$	&	$-0.07$	&	$0.141(35)$&	$-0.004(1)	$	&	$	-	$   & $-$	\\
$Z=70$	&	$\mathbf{170}$	&	$\mathbf{100}$	&	$5.285(6)$	&	$3.15(1)$	&	$0.924(1)$	&	$-87.3(1)$	&	$	6.27(8)	$	&	$0.01$	&			&	$0.153(38)$&	$-0.004(1)	$	&	$-0.0003$    & $0.3$	\\				
		&	$\mathbf{171}$	&	$\mathbf{101}$	&	$5.291(6)$	&	$3.15(1)$	&	$0.924(1)$	&	$-88.2(1)$	&	$	6.34(8)	$	&	$0.02$	&			&	\\										
		&	$\mathbf{172}$	&	$\mathbf{102}$	&	$5.300(6)$	&	$3.15(1)$	&	$0.924(1)$	&	$-89.1(1)$	&	$	6.40(8)	$	&	$0.03$	&			&	$0.174(40)$&	$-0.005(1)	$	&	$-0.0008$    & $0.4$	\\				
		&	$\mathbf{173}$	&	$\mathbf{103}$	&	$5.305(6)$	&	$3.15(1)$	&	$0.924(1)$	&	$-90.0(1)$	&	$	6.46(8)	$	&	$0.03$	&			&	\\										
		&	$\mathbf{174}$	&	$\mathbf{104}$	&	$5.311(6)$	&	$3.15(1)$	&	$0.924(1)$	&	$-91.0(1)$	&	$	6.52(8)	$	&	$0.04$	&			&	$0.202(51)$&	$-0.005(1)	$	&	$-0.0016$    & $0.5$	\\				
		&	$\mathbf{176}$	&	$\mathbf{106}$	&	$5.322(6)$	&	$3.14(1)$	&	$0.924(1)$	&	$-92.8(1)$	&	$	6.64(9)	$	&	$0.05$	&			&	$0.215(67)$&	$-0.006(2)	$	&	$-0.0019$    & $0.4$	\\				
\hline																																	
Pb		&	$202$			&	$120$			&	$5.471(2)$	&	$4.70(2)$	&	$0.897(2)$	&	$-101.7(2)$	&	$	10.3(1)	$	&	$-	$	&	$-0.10$	&	\\									
$Z=82$	&	$\mathbf{204}$	&	$\mathbf{122}$	&	$5.480(1)$	&	$4.69(2)$	&	$0.897(2)$	&	$-103.5(2)$	&	$	10.4(1)	$	&	$0.01$	&			&	$0.172(44)$&	$-0.006(2)	$	&	$	-	$ & $-$	\\				
		&	$205$			&	$123$			&	$5.483(2)$	&	$4.69(2)$	&	$0.897(2)$	&	$-104.4(2)$	&	$	10.5(1)	$	&	$0.02$	&			&	\\											
		&	$\mathbf{206}$	&	$\mathbf{124}$	&	$5.490(1)$	&	$4.69(2)$	&	$0.897(2)$	&	$-105.3(2)$	&	$	10.6(1)	$	&	$0.02$	&			&	$0.184(46)$&	$-0.007(2)	$	&	$-0.0004$ & $0.5$	\\					
		&	$\mathbf{207}$	&	$\mathbf{125}$	&	$5.494(1)$	&	$4.69(2)$	&	$0.897(2)$	&	$-106.2(2)$	&	$	10.7(1)	$	&	$0.03$	&			&	\\											
		&	$\mathbf{208}$	&	$\mathbf{126}$	&	$5.501(1)$	&	$4.69(2)$	&	$0.897(2)$	&	$-107.1(2)$	&	$	10.8(1)	$	&	$0.03$	&			&	$0.200(50)$&	$-0.007(2)	$	&	$-0.0010$ & $0.6$	\\					
\hline																																	
Fr		&	$207$			&	$120$			&	$5.57(2)$	&	$5.66(3)$	&	$0.884(2)$	&	$-100.0(2)$	&	$	11.5(1)	$	&	$-	$	&	$-0.11$	&	\\								
$Z=87$	&	$208$			&	$121$			&	$5.57(2)$	&	$5.66(3)$	&	$0.884(2)$	&	$-100.9(2)$	&	$	11.6(1)	$	&	$0.01$	&			&	\\										
		&	$209$			&	$122$			&	$5.58(2)$	&	$5.65(3)$	&	$0.884(2)$	&	$-101.7(2)$	&	$	11.7(1)	$	&	$0.01$	&			&	$0.121(36)$&	$-0.005(1)	$	&	$	-	$   & $-$	\\			
		&	$210$			&	$123$			&	$5.58(2)$	&	$5.65(3)$	&	$0.884(2)$	&	$-102.6(2)$	&	$	11.8(1)	$	&	$0.02$	&			&	\\										
		&	$211$			&	$124$			&	$5.59(2)$	&	$5.65(3)$	&	$0.884(2)$	&	$-103.5(2)$	&	$	11.9(1)	$	&	$0.02$	&			&	$0.132(38)$&	$-0.005(2)	$	&	$-0.0004$ & $0.5$	\\				
		&	$212$			&	$125$			&	$5.59(2)$	&	$5.65(3)$	&	$0.885(2)$	&	$-104.4(2)$	&	$	12.0(1)	$	&	$0.03$	&			&	\\										
		&	$213$			&	$126$			&	$5.60(2)$	&	$5.64(3)$	&	$0.885(2)$	&	$-105.3(2)$	&	$	12.1(1)	$	&	$0.03$	&			&	$0.146(42)$&	$-0.006(2)	$	&	$-0.0010$ & $0.5$	\\				
		&	$220$			&	$133$			&	$5.67(2)$	&	$5.61(3)$	&	$0.885(2)$	&	$-111.5(2)$	&	$	12.7(1)	$	&	$0.07$	&			&	\\										
		&	$221$			&	$134$			&	$5.68(2)$	&	$5.60(3)$	&	$0.885(2)$	&	$-112.4(2)$	&	$	12.8(1)	$	&	$0.08$	&			&	$0.206(53)$&	$-0.008(2)	$	&	$-0.0032$ & $0.6$	\\				
		&	$222$			&	$135$			&	$5.69(2)$	&	$5.60(3)$	&	$0.885(2)$	&	$-113.3(2)$	&	$	12.9(1)	$	&	$0.08$	&			&	\\										
		&	$223$			&	$136$			&	$5.70(2)$	&	$5.59(3)$	&	$0.885(2)$	&	$-114.2(2)$	&	$	13.0(1)	$	&	$0.09$	&			&	\\										
		&	$224$			&	$137$			&	$5.71(2)$	&	$5.59(3)$	&	$0.885(2)$	&	$-115.1(2)$	&	$	13.1(1)	$	&	$0.09$	&			&	\\										
		&	$225$			&	$138$			&	$5.71(2)$	&	$5.59(3)$	&	$0.885(2)$	&	$-116.0(2)$	&	$	13.2(1)	$	&	$0.09$	&			&	\\										
		&	$226$			&	$139$			&	$5.72(2)$	&	$5.58(3)$	&	$0.885(2)$	&	$-116.9(2)$	&	$	13.3(1)	$	&	$0.10$	&			&	\\										
		&	$227$			&	$140$			&	$5.73(2)$	&	$5.58(3)$	&	$0.885(2)$	&	$-117.8(2)$	&	$	13.4(1)	$	&	$0.10$	&			&	\\										
		&	$228$			&	$141$			&	$5.74(2)$	&	$5.57(3)$	&	$0.885(2)$	&	$-118.6(2)$	&	$	13.5(1)	$	&	$0.11$	&			&	\\										
\hline																																	
Ra		&	$208$			&	$120$			&	$5.59(2)$	&	$5.89(3)$	&	$0.882(2)$	&	$-99.6(2)$	&	$	11.8(1)	$	&	$-	$	&	$-0.11$	&	\\							
$Z=88$	&	$209$			&	$121$			&	$5.59(2)$	&	$5.89(3)$	&	$0.882(2)$	&	$-100.5(2)$	&	$	11.9(1)	$	&	$0.01$	&			&	\\										
		&	$210$			&	$122$			&	$5.59(2)$	&	$5.88(3)$	&	$0.882(2)$	&	$-101.4(2)$	&	$	12.0(1)	$	&	$0.01$	&			&	$0.111(34)$&	$-0.005(1)	$	&	$	-	$ & $-$	\\			
		&	$211$			&	$123$			&	$5.59(2)$	&	$5.88(3)$	&	$0.882(2)$	&	$-102.3(2)$	&	$	12.1(1)	$	&	$0.02$	&			&	\\										
		&	$212$			&	$124$			&	$5.60(2)$	&	$5.88(3)$	&	$0.882(2)$	&	$-103.1(2)$	&	$	12.2(1)	$	&	$0.02$	&			&	$0.123(37)$&	$-0.005(2)	$	&	$	-0.0005$ & $0.5$	\\				
		&	$213$			&	$125$			&	$5.60(2)$	&	$5.88(3)$	&	$0.882(2)$	&	$-104.0(2)$	&	$	12.3(1)	$	&	$0.03$	&			&	\\										
		&	$214$			&	$126$			&	$5.61(2)$	&	$5.87(3)$	&	$0.882(2)$	&	$-104.9(2)$	&	$	12.4(1)	$	&	$0.03$	&			&	$0.136(40)$&	$-0.006(2)	$	&	$	-0.0010$ & $0.5$	\\				
		&	$220$			&	$132$			&	$5.67(2)$	&	$5.84(3)$	&	$0.882(2)$	&	$-110.2(2)$	&	$	12.9(1)	$	&	$0.07$	&			&	$0.181(49)$&	$-0.008(2)	$	&	$	-0.0028$ & $0.6$	\\				
		&	$221$			&	$133$			&	$5.68(2)$	&	$5.83(3)$	&	$0.882(2)$	&	$-111.1(2)$	&	$	13.0(1)	$	&	$0.07$	&			&	\\										
		&	$222$			&	$134$			&	$5.69(2)$	&	$5.83(3)$	&	$0.882(2)$	&	$-112.0(2)$	&	$	13.1(1)	$	&	$0.08$	&			&	$0.195(52)$&	$-0.008(2)	$	&	$	-0.0033$ & $0.6$	\\				
		&	$223$			&	$135$			&	$5.70(3)$	&	$5.82(3)$	&	$0.882(2)$	&	$-112.9(2)$	&	$	13.2(1)	$	&	$0.08$	&			&	\\										
		&	$224$			&	$136$			&	$5.70(3)$	&	$5.82(3)$	&	$0.882(2)$	&	$-113.8(2)$	&	$	13.3(1)	$	&	$0.09$	&			&	\\										
		&	$225$			&	$137$			&	$5.72(3)$	&	$5.82(3)$	&	$0.882(2)$	&	$-114.7(2)$	&	$	13.4(1)	$	&	$0.09$	&			&	\\										
		&	$226$			&	$138$			&	$5.72(3)$	&	$5.81(3)$	&	$0.882(2)$	&	$-115.6(2)$	&	$	13.5(1)	$	&	$0.10$	&			&	\\										
		&	$227$			&	$139$			&	$5.73(3)$	&	$5.81(3)$	&	$0.882(2)$	&	$-116.4(2)$	&	$	13.6(2)	$	&	$0.10$	&			&	\\										
		&	$228$			&	$140$			&	$5.74(3)$	&	$5.80(3)$	&	$0.882(2)$	&	$-117.3(2)$	&	$	13.7(2)	$	&	$0.10$	&			&	\\										
		&	$229$			&	$141$			&	$5.75(3)$	&	$5.80(3)$	&	$0.882(2)$	&	$-118.2(2)$	&	$	13.8(2)	$	&	$0.11$	&			&	\\										
		&	$230$			&	$142$			&	$5.76(3)$	&	$5.79(3)$	&	$0.883(2)$	&	$-119.1(2)$	&	$	13.9(2)	$	&	$0.11$	&			&	\\										
		&	$232$			&	$144$			&	$5.77(4)$	&	$5.78(3)$	&	$0.883(2)$	&	$-120.9(2)$	&	$	14.1(2)	$	&	$0.12$	&			&	\\										
\hline \hline

\end{longtable*} 

\begin{table}[tb]
\caption{Relative contribution of the neutron-skin effect to the total PNC amplitude or a ratio of PNC amplitudes in Yb isotopes used in experiment \cite{antypas_isotopic_2019}, evaluated according to Eqs.\,\eqref{eq:dEE_general}, \eqref{eq:chi} and \eqref{eq:delta_nsk}. We compute the difference between neutron and proton rms radii $\Delta r_{np}$ from \cite{brown_calculations_2009} and nuclear charge radii $r_p$ from \cite{angeli_table_2013}. The ratio $\chi$ of neutron skin contribution to proton contribution in an isotopic ratio $\mathcal{R}$ is defined in Eq. \eqref{eq:chi}.}\label{tab:Yb}
\begin{tabular}{p{0.5cm} p{0.4cm} p{0.6cm} p{1.3cm} p{1.3cm} p{1.5cm} p{1.4cm} p{0.5cm}}
 	&	$Z$	&	$A$	& $r_p$, fm & $\Delta r_{np}$, fm & $\frac{\delta E^\mathrm{n.s.}_\mathrm{PNC}}{E_\mathrm{PNC}}$	& $\Delta_\mathrm{n.s.}/\mathcal{R}$ & $\chi$	\\
\hline															
Yb	&	70	&	170	&	$5.285(6)$ &	$0.153(38)$	&	$-0.004(1)$ & $-$   &    $-$\\
	&		&	172	&	$5.300(6)$ &	$0.174(40)$	&	$-0.005(1)$ & $-0.0005$ &  $0.5$\\
    &		&	174	&	$5.311(6)$ &	$0.202(51)$	&	$-0.005(1)$ & $-0.0013$ & $0.6$\\
    &		&	176	&	$5.322(6)$ &	$0.215(67)$	&	$-0.006(2)$ & $-0.0016$ & $0.5$\\
\hline\hline										
\end{tabular}
\end{table}

In Ref.\,\cite{antypas_isotopic_2019} the PNC effects were measured in four even isotopes of Yb. Table \ref{tab:Yb} details the size of the neutron-skin contribution to the PNC amplitude in individual Yb isotopes and their ratios, according to neutron skin values from \cite{brown_calculations_2009} and Eqs.\,\eqref{eq:K} and \eqref{eq:delta_nsk}. The listed values indicate that it may be possible to detect the neutron skin contribution in an isotopic ratio of Yb if $0.1\%$ level of accuracy is reached.

In order to obtain useful experimental information regarding the new physics couplings or the behavior of neutron skins, one would typically need to know the ratio $\mathcal{R}$ with $\sim 0.1\%$ accuracy. Pairs of isotopes with large $\Delta N$, e.g. $^{209}$Fr and $^{221}$Fr or $^{144}$Sm and $^{154}$Sm, offer an estimated contribution of neutron skin in the isotopic ratio greater than $0.1\%$ and one can expect that a pair of the lightest and the heaviest isotope in Fr or Ra would produce even greater neutron skin contribution.

\section{Conclusion}\label{sec:conclusion}
In this work, we provided numerical evaluation of the dependence of spin-independent PNC effect on new physics and neutron skin effects for atoms of current experimental interest. On the basis of the data presented above, one can select the most suitable isotopes and pairs of isotopes for further PNC experiments. We estimated the size of neutron skin corrections for both single isotopes and isotopic ratios. If a 0.1\% accuracy in the isotopic ratio measurement is reached, neutron skin effects can be detected in any of the eight elements presented in Table \ref{tab:results}, given that isotopes with the largest difference in neutron number $\Delta N$ are chosen. It was discovered that the uncertainty in Fermi-model diffuseness $z$ (and, therefore, in the model of nuclear density in general) plays a crucial role in the extraction of the neutron skin from the single-isotope experiments, which poses a challenge for improvement of nuclear models.

Isotopic ratios give an opportunity to probe new physics corrections to the standard model proton couplings in low-energy experiments, and, in contrast to single-isotope experiments, without the need to calculate many-body electronic factor of the PNC amplitude. If the neutron skin uncertainties cancel out \cite{brown_calculations_2009}, isotopic ratios can constitute a valuable tool for the search of a list of new physics scenarios, such as new $Z'$ bosons \cite{davoudiasl_muon_2014,antypas_isotopic_2019,dzuba_probing_2017,safronova_search_2018}. We found that if a large contribution of neutron skin is to be avoided, it may be beneficial to choose a pair of isotopes with a large $\Delta N$ of one of the lighter elements. On the other hand, for experiments aiming at detecting the neutron skin, heavier elements such as Yb, Pb, Fr and Ra possess a pronounced neutron skin contribution to the PNC effects.

\section{Acknowledgements}
This work is supported by the Gutenberg Research College fellowship and Australian Research Council. The authors would like to thank Pierre Capel, Sonia Bacca and Vladimir Zelevinsky for valuable discussions.

\section*{Appendix}
To obtain the radial dependence of one-electron $s$ and $p_{1/2}$ wave functions, we assume the nucleus to be a sharp-edge charged sphere with radius $R_p$. In this approximation, the electron experiences the potential of a 3D harmonic oscillator inside the nucleus: 
\begin{equation}
V(r)=\frac{-Z e^{2}}{R_p} \left(\frac{3}{2} - \frac{r^{2}}{2 R_p^{2}} \right)\ .
\end{equation}
Thus, wave functions inside the nucleus are those of a quantum relativistic harmonic oscillator. Outside the nucleus, they should be relativistic Coulomb wave functions of a point charge. The harmonic-oscillator wave functions should be properly normalized to match Coulomb wave functions (presented, for example, in \cite{khriplovich_parity_1991}) at the edge of the nucleus. The $s$ and $p_{1/2}$ wave functions inside the nucleus can be thus written as \cite{flambaum_nuclear_2002,khriplovich_parity_1991}:
\begin{equation}
\psi_{s_{1/2}}= \pten{F_{s} \Omega_{s}}{i G_{s} \Omega_{p_{1/2}}} ,
\end{equation}
\begin{equation}
\psi_{p_{1/2}}=\pten{-\frac{A_{p}}{A_{s}}\;G_{s} \Omega_{p_{1/2}}}{i \frac{A_{p}}{A_{s}}\;F_{s} \Omega_{s}},
\end{equation}
where $\Omega_{s}$ and $\Omega_{p_{1/2}}$ are spherical functions with spin (eigenfunctions of total angular momentum operators $\hat{J_z}$ and $\hat{J^2}$), normalized as $\int d\phi \int d\theta \Omega^\dagger\Omega = 1$. Denoting $x=r/R_p$, one can write a Taylor expansion of the oscillator radial functions inside the nucleus up to $Z^2\alpha^2$ order \cite{flambaum_nuclear_2002,khriplovich_parity_1991}:
\begin{equation}
F_s(x)=A_s\left[1-\frac{3}{8}Z^2\alpha^2x^2\left(1-\frac{4}{15}x^2+\frac{1}{45}x^4\right) \right]\ ,
\end{equation}
\begin{align}
G_{s}(x) = & -\frac{1}{2} A_{s} Z \alpha x \nonumber \\ &\times \left \lbrack 1 - \frac{1}{5} x^2 -\frac{9}{40} Z^{2}\alpha^{2}  x^2 \left( 1 - \frac{3}{7} x^2 + \frac{4}{81}x^4 \right)  \right \rbrack\ .
\end{align}
Here the normalization constants $A_s$ and $A_p$ can be approximated \cite{khriplovich_parity_1991} as
\begin{align}
A_{s} = \frac{2}{(z_i+1)^{1/2}}\frac{2 (\frac{a_B}{2ZR_p})^{1-\gamma}}{\Gamma(2 \gamma + 1)} \left( \frac{Z}{a_B^{3}} \right )^{1/2}\nonumber \\ \times \left( \frac{I}{\mathrm{Ry}} \right )^{3/4} \left( 1 - \frac{1}{40} Z^{2}\alpha^{2} \right)\ ,
\end{align}
\begin{align}
A_{p} = \frac{Z\alpha}{(z_i+1)^{1/2}} \frac{2 (\frac{a_B}{2ZR_p})^{1-\gamma}}{\Gamma(2 \gamma + 1)} \left( \frac{Z}{a_B^{3}} \right )^{1/2} \nonumber\\ \times \left( \frac{I}{\mathrm{Ry}} \right )^{3/4}  \left( 1 + \frac{9}{40} Z^{2}\alpha^{2} \right)\ ,
\end{align}
with  $\gamma = \sqrt{1 - Z^{2}\alpha^{2}}$, $I$ being the ionization energy of the valence electron and the Rydberg constant $\mathrm{Ry}=e^{2}/(2 a_B)$. The Bohr radius is denoted as $a_B$. 

Let us expand the expression used in calculating the matrix element \eqref{eq:ME1} with the wave functions defined above:
\begin{equation}
\psi_{p_{1/2}}^\dagger\gamma_5\psi_s = -i\frac{A_p}{A_s}\left(G_s^2\Omega_{p_{1/2}}^\dagger\Omega_{p_{1/2}}+F_s^2\Omega_s^\dagger\Omega_s\right)
\end{equation}
If we denote $f_s=F_s/A_s$ and $g_s=G_s/A_s$ and take into account the normalization of spherical functions, then 
\begin{align*}
\int\psi_{p_{1/2}}^\dagger&\gamma_5\psi_s d^3r=-i\frac{A_p}{A_s}\int_0^\infty\left(G_s^2+F_s^2\right)dr\\&=-i A_pA_s\int_0^\infty (g_s^2+f_s^2)dr \\ &\equiv \mathcal{A}_{ps} \left(\frac{2ZR_p}{a_B}\right)^{2\gamma-2}\int_0^\infty f_{ps}(r)dr\ .
\end{align*}
Here $f_{ps}(r)$, such that $f_{ps}(0)=1$, contains the variation of wave functions density inside the nucleus, the normalization coefficient $\mathcal{N}\equiv\left(2ZR_p/a_B\right)^{2\gamma-2}$ shows the dependence of the expression on equivalent charge radius and $\mathcal{A}_{ps}$ consists of all remaining constants. We use above definitions to calculate the matrix element \eqref{eq:ME1} in the main text.

\end{document}